\documentclass[aps,superscriptaddress,showpacs,nofootinbib]{revtex4}

\usepackage{epsfig,verbatim} 
\input{epsf}

\begin{document}

\title{Updating the Phase Diagram of the Gross-Neveu Model in $2+1$
  Dimensions}

\author{Jean-Lo\"{\i}c Kneur} \email{kneur@lpta.univ-montp2.fr}
\affiliation{Laboratoire de Physique Th\'{e}orique et Astroparticules - CNRS -
  UMR 5207 Universit\'{e} Montpellier II, France}

\author{Marcus Benghi Pinto} \email{marcus@fsc.ufsc.br}
\affiliation{Departamento de F\'{\i}sica, Universidade Federal de Santa
  Catarina, 88040-900 Florian\'{o}polis, Santa Catarina, Brazil}

\author{Rudnei O. Ramos} \email{rudnei@uerj.br} \affiliation{Departamento de
  F\'{\i}sica Te\'{o}rica, Universidade do Estado do Rio de Janeiro, 20550-013
  Rio de Janeiro, RJ, Brazil}

\author{Ederson Staudt} \email{ederson@fsc.ufsc.br} \affiliation{Departamento
  de F\'{\i}sica, Universidade Federal de Santa Catarina, 88040-900
  Florian\'{o}polis, Santa Catarina, Brazil}

\begin{abstract}
  The method of optimized perturbation theory (OPT) is used to study the phase
  diagram of the massless Gross-Neveu model in $2+1$ dimensions. In the
  temperature and chemical potential plane, our results give strong support to
  the existence of a tricritical point and line of first order phase
  transition, previously only suspected to exist from extensive lattice Monte
  Carlo simulations. In addition of presenting these results we discuss how
  the OPT can be implemented in conjunction with the Landau expansion in order
  to determine all the relevant critical quantities.
  
  \medskip
  
  \medskip

\noindent
keywords: chiral phase transition, Gross-Neveu model, four-Fermi models,
nonperturbative thermodynamics

\end{abstract}


\maketitle

\section{Introduction}

The use of exactly solvable models and methods leading to analytical solutions
has always been important in physics in general. Models of particular interest
are those that display some physical correspondence to the physics of quantum
chromodynamics (QCD) or that may give insights to the difficult task, given
its nonperturbative nature, of analyzing the phase structure at finite
temperatures and densities. This is also the kind of situation, regarding
phase transitions in a hot and dense medium, that is expected to be realized
in the early universe and in the recent experiments involving heavy-ion
collisions.  The use of appropriate techniques and models in the context of
quantum field theory that can increase our knowledge on this kind of phenomena
are then timely and important.  One such phenomenon of topical interest is the
study of the chiral symmetry restoration (CSR) in QCD.  However, due to the
intrinsic difficulty of approaching directly this study in the context of QCD
itself, other more tractable models have been extensively used.  Among them,
the simplest fermionic model displaying many properties common to QCD is the
Gross-Neveu (GN) model \cite{GN}.  {}For instance, in 2+1 dimensions it can
exhibit a discrete spontaneous chiral symmetry breaking (CSB) that happens for
a critical coupling at zero temperature and density, while its spectrum can
contain excitations like baryons as well as mesons (which are composite
fermionic states). Even though it is a quartic interacting fermion model, thus
nonrenormalizable (in 2+1 dimensions) in ordinary perturbation theory, it is
renormalizable in a $1/N$ expansion (where $N$ denotes the number of fermion
species) and it is exactly soluble in the large-$N$ limit \cite{Park}.  These
properties make this effective model one of the simplest and most useful
frameworks for understanding some of the important physical features displayed
by QCD.

{}For the reasons above, the GN model is a perfect laboratory for studying
symmetry aspects in a hot and dense medium and outside the QCD realm, the
model by its own and its extensions may also have applications in condensed
matter physics in the understanding of the properties of fermion liquids and
high-$T_c$ superconductivity models \cite{condmatter}.  Besides being studied
analytically mostly with the large-$N$ expansion technique, there are a large
number of works studying the GN model numerically from the lattice Monte-Carlo
perspective \cite{lattice1,lattice2} (for an introduction to lattice QCD, its
problems in implementation and applications to effective models, like the GN
one, see \cite{latticeQCD} and references therein).  Among the interesting
thermodynamical properties of the GN model demonstrated by the large-$N$
technique is the existence of a line of second order CSR in the temperature
and chemical potential ($T,\mu$) plane, starting at a value ($T_c,0$) and
ending, abruptly, at a first order transition point ($0,\mu_c$)
\cite{Park,klimenko}.  While this phase structure was verified by lattice
simulations with finite number of fermion species, $N$, in Ref.
\cite{lattice2} some indication was given for the existence of a first order
transition line starting at a tricritical point ($T_{\rm tcr},\mu_{\rm tcr}$)
and ending at $\sim (0,\mu_c)$. However, the precise position of this
tricritical point in the phase diagram could not be determined.  Even within
the $1/N$ expansion, there are no results available on the phase structure of
the 2+1 dimensional GN model beyond the leading order, due to the appearance
of complicate integrals (note however that there are results up to ${\cal
  O}(1/N^2)$ for the critical exponents \cite{Gracey} of the bulk transition).
It then becomes an important and interesting issue to investigate whether
there is indeed a tricritical point for the GN model at finite $N$ in 2+1
dimensions and by which method we would be able to localize it in a robust
form. By doing so, we expect to have a much better understanding of the chiral
symmetry restoration problem at finite $T$ and $\mu$ not only in this model
but also to have eventually a better knowledge of a similar behavior that may
happen in QCD or even in analog condensed matter systems.  In this work we
study the 2+1 dimensional GN model beyond the large-$N$ limit and, to obtain
its phase diagram at finite $T$ and $\mu$, we make use of the method of the
optimized perturbation theory (OPT) (also known as linear $\delta$ expansion)
\cite{LDE}. Here, we discuss how this method can be used to extract useful
information, regarding critical quantities, from the Landau expansion of the
free energy density.

This paper is organized as follows. In the next section we introduce the
model, presenting the OPT free energy density. In Sec. III, we discuss
optimization issues related to the Landau expansion of the free energy
density. Our results for the $(T,\mu)$ phase diagram as well as for the
critical quantities are presented in Sec. IV, showing the existence of
tricritical point for finite values of $N$.  Our conclusions and perspectives
for future applications are discussed in Sec. V.

\section{The Interpolated Gross-Neveu Model}

The GN model is described by a Lagrangian density for a $N$ component fermion
field $\psi_j$ ($j=1,\ldots,N$) with local quartic fermion interaction which
is defined by \cite{GN} (in this work all our expressions are defined in
Euclidean space)

\begin{equation}
{\cal L} =
\bar{\psi}_{j}   \not\!\partial \psi_{j} - 
\frac {\lambda}{2 N} ({\bar \psi_j} \psi_j)^2\;,
\label{gn}
\end{equation}
where, in 2+1 dimensions, the fermion fields are treated as four-component
Dirac spinors and the $\gamma_\mu$ matrices are defined in the $4 \times 4$
representation. In this way, Eq. (\ref{gn}) is invariant under the discrete
chiral transformations $\psi_j \to \gamma_5 \psi_j$ and $\bar{\psi}_j \to -
\bar{\psi}_j \gamma_5$.  Though it also has a continuous flavor symmetry we
will only be interested in the case of discrete symmetry, recalling a general
no-go theorem in 2+1 dimensions that forbids spontaneous symmetry breaking of
a continuous symmetry at any finite temperature \cite{coleman}.

The implementation of the OPT within this model is reviewed in a previous
application \cite{LPR} to the 1+1 dimensional case.  In practice, one
considers the original theory, Eq.  (\ref{gn}), adding a quadratic (in the
fermion field) term $(1-\delta) \eta {\bar \psi_j} \psi_j$ where $\eta$ is an
arbitrary mass parameter and $\delta$ is a bookkeeping parameter which labels
the order at which the OPT is being performed.  At the same time, $\lambda \to
\delta \lambda$. Also, as conventional, we rewrite the quartic term by
introducing an auxiliary scalar field $\sigma$. Then, Eq. (\ref{gn}) acquires
its OPT form \cite {LPR}

\begin{equation}
{\cal L}_{\delta} =
\bar{\psi}_{j}  \not\!\partial  \psi_{j} +
\delta \sigma {\bar \psi_j} \psi_j +  (1-\delta)\eta {\bar \psi_j} \psi_j
+ \frac {\delta N }{2 \lambda } \sigma^2 \;.
\label{GNdelta}
\end{equation}
The OPT then describes an interpolated theory with a fermion propagator $[\not
\! P + (1-\delta) \eta +\delta \sigma_c ]^{-1}$, Yukawa vertex $\delta$ and
$\sigma$ propagator $\lambda/(N \delta)$.  Any quantity $\Phi^{(k)}$ is then
evaluated to an order-$\delta^k$ using these new {}Feynman rules.  As shown in
the many previous applications using this method \cite{moreOPT}, it operates
for any $N$ and is free from infrared divergences.  Nonperturbative results
for the original ($\eta$-independent) theory are obtained by optimizing
$\Phi^{(k)}$ with respect to $\eta$ at $\delta=1$. One standard and largely
adopted optimization procedure is the variational criterion known as the
Principle of Minimal Sensitivity (PMS) which fixes $\eta$ by requiring, order
by order, that $ d \Phi^{(k)} /d \eta =0$ \cite{PMS}.

{}From Eq. (\ref{GNdelta}), we are interested in evaluating the effective
potential for the auxiliary scalar background field, $V_{\rm eff} (\sigma_c)$,
obtained once all fermion fields and fluctuations around the scalar background
field are integrated out \cite {Park}. {}From this quantity all the
thermodynamics for the model can be derived.  In particular, CSB is signaled
by a non-vanishing vacuum expectation value for $\sigma$, ${\bar \sigma}_c
\equiv \langle \sigma \rangle$, which is a minimum of $V_{\rm eff}
(\sigma_c)$.

\begin{figure}[htb]
  \vspace{0.5cm}
  \centerline{\epsfig{figure=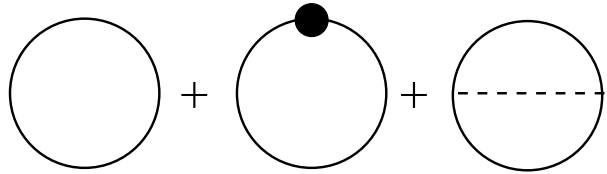,angle=0,width=8cm}}
\caption[]{Feynman diagrams contributing to $V_{\rm eff}$ to
  ${\cal O}(\delta^1)$.  The continuous lines are (the $\delta$ independent)
  fermion propagators, the black dot is an insertion of $\delta (\sigma_c -
  \eta)$, while the dashed line represents the $\sigma$ propagator.
\label{diagrams}}
\end{figure}

In {}Fig. \ref{diagrams} we show all the {}Feynman diagrams contributing to
$V_{\rm eff}(\sigma_c)$ up to ${\cal O}(\delta)$.  Note that already at this
first order in the OPT a correction that goes beyond the large-$N$ expansion
is included. The first two diagrams in {}Fig. \ref{diagrams} are ${\cal
  O}(1/N^0)$, while the third one contains an exchange type of self-energy as
insertion and brings the first ${\cal O} (1/N)$ correction to $V_{\rm
  eff}(\sigma_c)$.
Explicitly, we have for all the terms shown in {}Fig. \ref{diagrams} the
expression


\begin{equation}
\frac{V_{\rm eff,\delta^1}(\sigma_c,\eta)}{N} =
\delta \frac{\sigma_c^2}{2\lambda} - \int_p^{(T)}
{\rm tr} \, \ln (\not \! P + \eta) 
 - \delta (\sigma_c - \eta)
\int_p^{(T)}{\rm tr}
\frac{1}{\not \! P + \eta}
+ \delta \frac{\lambda}{2 N}
\int_p^{(T)} \int_q^{(T)}
{\rm tr} \, \frac{1}{(\not \! P + \eta)(\not \! Q + \eta)}\;,
\label{Veff}
\end{equation}
where the Euclidean momenta are given by $P=(P_0,{\bf p})$ and $P_0 = (2
n+1)\pi T - i \mu$, with $n=0, \pm 1, \pm 2, \ldots$. In the above equation we
have used the compact notation:

\[
\int_p^{(T)} = T\sum_{n=-\infty}^{+\infty} \int \frac{d^{d-1} p}{(2
  \pi)^{d-1}},
\]
with all momentum space integrals done with dimensional regularization, $d=3-
\epsilon$, while renormalization can be implemented in the modified minimal
subtraction scheme ($\overline{\rm MS}$). All the contributions considered in
Eq. (\ref {Veff}) are actually finite in dimensional regularization.  In Ref.
\cite{long} we have considered the free energy up to order-$\delta^2$, at
$T=0$ and $\mu=0$, where the first divergences show up and we refer the
interested reader to that work for a detailed account on the renormalization
problem of the 2+1 dimensional GN model in the OPT.  Let us only emphasize
here that though the model is not perturbatively renormalizable, to ${\cal
  O}(\delta^2)$ and within dimensional regularization, we only need standard
(i.e. mass, wave-function, etc) renormalization counterterms to render the
effective potential calculation finite \cite{long}.

\section{Landau's Expansion and the Optimization Procedure}

The effective potential, $V_{\rm eff}(\sigma_c)$, is a natural quantity to be
optimized in the spirit of the PMS procedure since it represents the system's
(Landau) free energy density, whose knowledge allows for a complete analysis
of all relevant thermodynamical properties.  Also, as explained in Ref.
\cite{LPR}, by optimizing the free energy we can guarantee that not only the
exact large-$N$ results are already obtained at leading order in the OPT, but
also at any subsequent order, provided that one stays within the $N \to
\infty$ limit (at first order in the OPT this corresponds to dropping the
third diagram in {}Fig. \ref{diagrams} and applying the optimization procedure
to the first two diagrams).  This shows that the OPT is convergent at any
order in the OPT when contrasted with the large-$N$ predictions. This result
is valid for any temperature or value of chemical potential and applies as
well to any number of dimensions (for other accounts on the convergence of the
OPT in the context of quantum field theory, please see \cite{conv} and
references therein). Let us consider the expression for $V_{\rm
  eff}(\sigma_c)$ at ${\cal O}(\delta)$ with $\lambda \to -\lambda$ in which
case chiral symmetry is broken at $T=0=\mu$. After taking the traces in Eq.
(\ref{Veff}), rearranging the terms and employing the redefinition of the
coupling, $\Lambda= \pi/|\lambda|$, one obtains

\begin{equation}
\frac{V_{{\rm eff},\delta^1}}{N} =-\delta
\frac {\sigma_c^2 \Lambda}{2 \pi} -
 2{\cal I}_0(\eta,\mu,T)+ 4 \delta\,
\eta\, (\eta-\sigma_c) \,{\cal I}_1(\eta,\mu,T)
- \delta  \frac{2 \pi}{N \Lambda} \eta^2 [{\cal I}_1(\eta,\mu,T)]^2 -
\delta \frac{2 \pi}{N \Lambda}[  {\cal I}_2(\eta,\mu,T) ]^2  \,\,,
\label{vlde1ddim}
\end{equation}
where

\begin{equation}
{\cal I}_0(\eta,\mu,T)=\int_p^{(T)} \ln \left( P^2 + \eta^2 \right)\,\,,\,\,
{\cal I}_1(\eta,\mu,T)=\int_p^{(T)}
\frac {1}{P^2 + \eta^2 }\,\,\,,\,\,\,
{\cal I}_2(\eta,\mu,T)=\int_p^{(T)}\frac {P_0}{P^2 + \eta^2 }\;.
\end{equation}

Setting $\delta=1$ and applying the PMS to Eq. (\ref{vlde1ddim}) we obtain a
self-consistent equation for the optimized mass


\begin{eqnarray}
{\bar \eta} = \sigma_c+ \frac {\pi}{N \Lambda} 
\left\{\eta {\cal I}_1(\eta,\mu,T)
+  {\cal I}_2(\eta,\mu,T){\cal I}_2^\prime(\eta,\mu,T) 
\left[{\cal I}_1(\eta,\mu,T) + 
\eta {\cal I}_1^\prime(\eta,\mu,T)\right]^{-1} \right\} 
\Bigr|_{\eta = {\bar \eta}} \,\,,
\label{pmsselfconsistent}
\end{eqnarray}
where the primes indicate derivatives with respect to $\eta$.  It can easily
be verified that one immediate consequence of Eq. (\ref{pmsselfconsistent}) is
that for $N \to \infty$ we have ${\bar \eta}=\sigma_c$, thus retrieving all
large-$N$ results, as mentioned in the previous paragraph.  All the details
concerning the Matsubara sums and dimensional regularized integrals can be
found in Ref. \cite{long}.  (Note that the sum over the Matsubara frequencies
and all space momentum integrals can be evaluated analytically at this first
$\delta$ order here considered).

In the case $\mu=0$, Eq.  (\ref{pmsselfconsistent}) factorizes in a nice way
which allows us to understand how the OPT resums the series producing
non-perturbative results.  Using ${\cal I}_2(\eta,0,T)=0$ (see Ref.
\cite{long}) we can write Eq. (\ref{pmsselfconsistent}) in the self-consistent
form:

\begin{eqnarray}
{\bar \eta} = \sigma_c + \frac {\pi}{N \Lambda}\bar{\eta}\:
{\cal I}_1(\bar{\eta},0,T) \,\,.
\label{pmseta}
\end{eqnarray}

Explicitly, ${\cal I}_1(\bar{\eta},\mu,T)$ is given by (from the Matsubara sum
and momentum integration)

\begin{equation}
{\cal I}_1(\bar{\eta},\mu,T)  = - \frac{T}{4\pi}
\left\{  \frac{|\bar{\eta}|}{T} +
\ln \left[ 1+ e^{-(|\bar{\eta}| - |\mu|)/T}
\right] +  \ln \left[1 + e^{-(|{\bar \eta}| + |\mu|)/T}\right] \right\}
\;.
\label{intp2}
\end{equation}

The resummation becomes even easier to be visualized if one notes that
$\Sigma_{\rm exc}(\bar{\eta},0,T) =\pi \bar {\eta}/ (N \Lambda)\: {\cal
  I}_1(\bar{\eta},0,T)$, where $\Sigma_{\rm exc}(\bar{\eta},0,T)$ represents
an exchange (rainbow, Fock) type of contribution so that, at $\mu=0$, one may
write ${\bar \eta} = \sigma_c +\Sigma_{\rm exc}(\bar{\eta},0,T)$.  Note that
prior to optimization, the terms contributing to $V_{\rm eff}$ to first order
are the ones given by Eq. (\ref{Veff}), represented by the diagrams shown in
{}Fig. \ref {diagrams}.  Upon optimization the diagrams are dressed by rainbow
type of self-energy contributions that originates from the first nontrivial
diagram represented by the third term in {}Fig. \ref {diagrams}.  This was
expected since at order-$\delta$ the perturbative OPT effective potential
receives information about this type of topology only.  The OPT will then
gather all powers of $1/N$ corresponding to the (rainbow) class of diagram.
In other words, the simple evaluation of a first topologically distinct
diagram is already able to bring non-perturbative information concerning that
type of contribution.  If one proceeds to order-$\delta^2$, information about
corrections to the scalar propagator as well as Yukawa vertex will enter the
perturbative effective potential. Then, the PMS will dress up these
perturbative contributions and so on.  Of course this rather simple OPT/PMS
procedure does not give all the truly non-perturbative and higher order
contributions, such as the genuine $1/N^2$ and higher order corrections, but
is just one particular class of graphs that are being resummed in this way. It
nevertheless gives a non trivial result that goes beyond the standard $1/N$
order.

Let us now consider the $T=\mu=0$ case. By optimizing $V_{\rm eff}(\sigma_c)$,
at ${\cal O}(\delta)$, it is very easy to show that one finds the optimum
value ${\bar \eta}= {\bar \sigma}_c {\cal F}(N)$, where ${\bar \sigma}_c$ is
the minimum of $V_{\rm eff}(\sigma_c)$, and ${\cal F}(N) =1 - 1/(4N)$ is a
characteristic function that will appear in all our expressions.  At this
value of $\eta$ and computing the extremum for the effective potential for
$\sigma_c$, we obtain that the order parameter is given by

\begin{equation}
{\bar \sigma}_c = \Lambda {\cal F}(N)^{-2}\;.
\label{sigmavev}
\end{equation}
Note that the scalar field vacuum expectation value ${\bar \sigma}_c$ sets the
natural scale for CSB. We could express all our other results in terms of
${\bar \sigma}_c$, but here we find more convenient to normalize all critical
quantities in terms of just $\Lambda$ (after all this is just a scaling
choice).  In Ref. \cite{long} the result for the critical temperature has been
obtained in a similar and completely analytic fashion producing the OPT result

\begin{equation}
T_c(\mu=0)= \Lambda/\left[ 2\ln2\,{\cal F}(N) \right]\;.
\label{Tc}
\end{equation} 
However, an interesting alternative is to obtain $T_c$ using the Landau
expansion for the free energy density, which is valid for small values of the
order parameter.  The Landau's expansion (for small $\sigma_c$) for $V_{\rm
  eff}$ can be expressed in the general form

\begin{equation}
V_{\rm eff}(\sigma_c,\mu,T) \simeq V_0 + \frac{1}{2} a(\mu,T)\: \sigma_c^2 +
\frac{1}{4} b(\mu,T) \:\sigma_c^4 + \frac{1}{6} c(\mu,T)\: \sigma_c^6 \;,
\label{landau}
\end{equation}
where $V_0$ is a constant (field independent) energy term.  Note that only
even powers of $\sigma_c$ are allowed due to the original chiral symmetry of
the model.  The coefficients $a,b$ and $c$ appearing in Eq. (\ref{landau}) can
be obtained, respectively, by a second, fourth and sixth derivative of the
free energy density expansion at $\sigma_c=0$. Higher order terms in the
expansion (\ref{landau}) can be verified to be much smaller than the first
order terms and can be consistently neglected.

Lets recall that the expression of the free energy in the form (\ref{landau})
is the simplest one able to exhibit a rich phase structure, allowing also for
studying tricritical phenomena.  In particular note that a tricritical point
can emerge whenever we have three phases coexisting simultaneously. {}For the
above potential, Eq. (\ref{landau}), one can have a second order phase
transition when the coefficient of the quadratic term vanishes ($a=0$) and $b
>0, c >0$.  A first order transition happens for the case of $b < 0, c > 0$.
The tricritical point occurs when both the quadratic and quartic coefficients
vanish, $a=b=0$ (with $c>0$). Thus, Eq. (\ref{landau}) offers a simple and
immediate way for analyzing the phase structure of our model.  {}For instance,
to obtain $T_c$ at $\mu=0$ one only needs to consider Eq.  (\ref {landau}) to
order-$\sigma_c^4$ with $b>0$ to assure that the potential is bounded from
below. Then, the solution of $a(0,T_c)=0$ sets the critical temperature.
However, in order to use Landau's expansion we must have $V_{\rm eff}$ in
terms of $\sigma_c$, $\mu$ and $T$ only (apart from $N$ and the scale
$\Lambda$, of course). In principle, this can be done by using the PMS
relation, Eq. (\ref {pmsselfconsistent}). As one can easily check the large
$N$ result, $T_c=\Lambda/(2\ln2)$ is quickly recovered by solving $a(0,T_c)=0$
with the latter obtained from Eq. (\ref {vlde1ddim}) with ${\bar \eta}=
\sigma_c$.  Now, at finite $N$, $\bar \eta$ depends on $\sigma_c$ in a highly
nonlinear way and the use of Landau's method in conjunction with the OPT-PMS
does not appear to be straightforward.  However, Eq.  (\ref{pmseta}) can then
be easily solved numerically by iteration.  Let us consider the first order
(in the iteration) approximate PMS solution (PMS1)

\begin{eqnarray}
{\bar \eta^{(1)}} \simeq \sigma_c + \frac {\pi}{N \Lambda}{\bar \eta}\: 
{\cal I}_1({\bar \eta},0,T)
 \Bigr|_{{\bar \eta}=\sigma_c} \,\,,
\label{pms1}
\end{eqnarray}
and to access the convergence of the solutions, let us also consider the next
approximation (PMS2) by doing ${\bar \eta} \to {\bar \eta}^{(1)}$ on the right
hand side of Eq. (\ref {pmseta}). In Table I we compare the results given by
PMS1 and PMS2 with the analytical relation $T_c=\Lambda/[2\ln2\,{\cal F}(N)]$,
Eq. (\ref{Tc}), for $\mu=0$.

\begin{table}[pt]
{\begin{tabular}{@{}c|c|c|c@{}} 
\hline 
$N$ & $T_c, Eq. (\ref{Tc})$ & $T_{c}^{\rm PMS1}$ & $T_c^{\rm PMS2}$ 
\\ \colrule
1 & 0.961797 & 0.979106  &   0.963579 \\
3 & 0.786925 & 0.786983   &  0.786925 \\
4 & 0.769438    & 0.769454 & 0.769437 \\
6 & 0.752710 & 0.752713  &  0.752710 \\
10 & 0.739844 & 0.739844  &  0.739844    \\
\hline 
\end{tabular}}
\caption{The critical temperature, $T_c$ (at $\mu=0$ and in units of 
$\Lambda$), obtained with the OPT 
in three alternative ways for different values of $N$. We have considered as 
much digits as we needed to establish eventual numerical differences. }
\end{table}
This shows that even the PMS1 approximation gives excellent results supporting
the possibility of applying the OPT-PMS in conjunction with Landau's
expansion. Especially, we notice that the results quickly converge for $N>4$
but even for the extreme case $N=1$ the agreement is rather good (higher than
$99\%$).

The reader should carefully note that we have written our results, in Table I,
up to seven digits just to compare the two numerical iterative optimization
procedures, PMS1 and PMS2 with the analytical solution, Eq. (\ref{Tc}). We
have considered as much digits as were needed in order to establish the
eventual numerical differences.  By no means this otherwise excessive number
of digits is intended to reflect the intrinsic accuracy of the OPT method at
lowest order. At the end of the next section we briefly comment on the
accuracy of the OPT based on the results at next order ${\cal O}(\delta^2)$.

\section{The Phase Diagram in the OPT}

Let us now present the phase diagram for the three dimensional model.  Since
the results for the cases $T=\mu=0$ and $T\ne 0,\mu=0$ have been discussed in
the previous section let us discuss the case $T=0$ and $\mu\neq 0$ for which
the relation ${\bar \eta}= {\bar \sigma}_c \, {\cal F}(N)$ is still valid (the
reader is referred to Ref. \cite {long} for details concerning the evaluation
of integrals and Matsubara's sums). In this case, CSR is found to happen at a
critical value for the chemical potential, $\mu_c$, which is obtained as
follows.  We first note that for $\mu=0$ (and $T=0$, as shown previously) the
effective potential has a minimum at the value of $\bar{\sigma}_c = \Lambda
{\cal F}(N)^{-2}$, as evaluated above.  {}For $\mu \neq 0$ the effective
potential acquires another minimum that gets degenerate with the first at a
value of chemical potential $\mu_c$, where $\bar{\sigma}_c=0$. At this point
we then find that $V_{\rm eff}(\bar\sigma_c=0, \mu=\mu_c, T=0) \equiv V_{\rm
  eff}(\bar\sigma_c, \mu=0,T=0)$, from where we get, after some algebra, that
the critical value for the chemical potential satisfies the equation

\begin{equation}
|\mu_c| =\frac{ \Lambda }{{\cal F}(N)}\,\left( 1+ \frac{3}{16N}\:
\frac{|\mu_c|}{\Lambda} \right)^{-1/3}\;.
\label{muc}
\end{equation}
Eq. (\ref{muc}) is a forth order equation for $\mu_c/\Lambda$, which is much
simpler to solve by iteration, converging very quickly since $\mu/\Lambda$ is
${\cal O}(1)$.  {}For example, at $N=3$, one finds $\mu_c \simeq 1.067
\Lambda$, which agrees with a full numerical evaluation of the complete $T$
and $\mu$ dependent expression (\ref{Veff}) at ${\cal O}(\delta)$ \cite{long}.
Since there is a potential barrier to overcome at the value of $\mu=\mu_c$,
the chiral transition here is discontinuous, thus first order.  Note also
that, for $N\to \infty$, Eq. (\ref{muc}) gives $\mu_c \to \Lambda$ and we
recover again a well known large-$N$ result \cite{Park}.

\begin{figure}[htb]
  \vspace{0.5cm}
  \centerline{\epsfig{figure=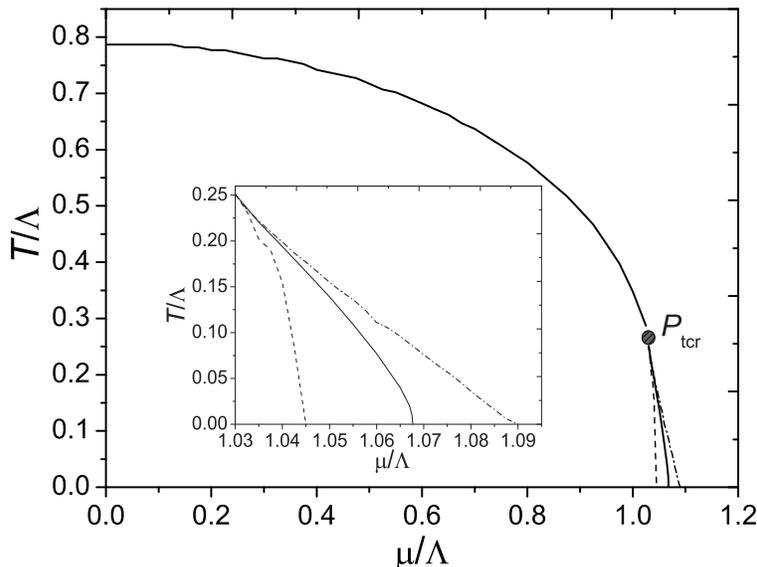,angle=0,width=10cm}} \vspace{-0.5cm}
\caption[]{\label{phase}
  The phase diagram for $N=3$ obtained within the OPT.  The position of the
  tricritical point is indicated by $P_{\rm tcr}$, where the second order
  transition line (above $P_{\rm tcr}$) turns into a first order one (below
  $P_{\rm tcr}$). CSB happens at the left of the continuous curve while CSR
  happens at the right hand side.  The dashed and dot-dashed lines are the
  metastable lines running along the first order line. The inner plot zooms
  into the region around the first order transition line. }
\end{figure}

{}Finally, we analyze the finite $T$ and $\mu$ effective potential. This is
done numerically for Eq. (\ref{Veff}) at ${\cal O}(\delta)$. {}For each value
of the temperature and chemical potential we apply the PMS optimization
condition to $V_{\rm eff}$ determining the optimized value of $\eta$,
$\bar{\eta}\equiv \bar{\eta}(\sigma_c,\mu,T)$.  This is inserted back in the
expression for $V_{\rm eff}$, from where we then obtain the regions for CSB
and CSR.  The result for the phase diagram, for $N=3$, is shown in {}Fig.
\ref{phase}.

According to the large-$N$ approximation the discrete chiral transition in the
GN model is of the second order everywhere except at $T=0$ and $\mu=\mu_c$.
The results obtained from the OPT, however, show the existence of a
tricritical point at $T_{\rm tcr} \simeq 0.25 \Lambda $ and $\mu_{\rm tcr}
\simeq 1.03 \Lambda$ (for $N=3$), represented by the black dot in {}Fig.
\ref{phase}.  Below this point the transition is of the first kind, while
above it is of the second kind. Metastable lines can be found at the left and
right of the first order critical curve on {}Fig. \ref{phase}, running very
close to it from the tricritical point down to the $\mu$ axis.  Any method
relying on determining this region, if not enough precise, is very likely to
miss it. This may then possibly explain why previous works, most notably
lattice Monte-Carlo simulations \cite{lattice2}, had difficulties to probe the
possible existence of a mixed phase.  It is possible to use the OPT-PMS in
conjunction with Landau's expansion in order to determine the tricritical
points for any finite values of $N$. In this case one can just reinstate the
chemical potential back in the last term of Eq. (\ref{pmseta}), iterate it to
the desired order, substitute the solution back in the expression for the
effective potential to then obtain the coefficients $a(\mu,T)$ and $b(\mu,T)$
in Eq. (\ref{landau}). Then by solving $a(\mu_{tcr},T_{tcr})=0$ and
$b(\mu_{tcr},T_{tcr})=0$ simultaneously, the tricritical points can easily be
extracted. As discussed above, direct numerical evaluation using the Eq.
(\ref{vlde1ddim}) and the PMS optimization give $T_{\rm tcr} \simeq 0.25
\Lambda $ and $\mu_{\rm tcr} \simeq 1.03 \Lambda$ for $N=3$. The Landau's
expansion, with for example the OPT-PMS2, gives $T_{\rm tcr} \simeq 0.24
\Lambda $ and $\mu_{\rm tcr} \simeq 1.03 \Lambda$. These results again compare
very well with the full numerical results used to obtain {}Fig. \ref{phase}.
In Table II we present all the relevant results for chiral symmetry
restoration for different values of $N$.  The quantities $T_c$ (at $\mu=0$),
$T_{tcr}$ and $\mu _{tcr}$ have been originally obtained in the present work
using the Landau expansion within the next order iteration PMS2, as described
above, while $\mu_c$ (at $T=0$) is given by the solution of Eq. (\ref{muc})
and the order parameter, ${\bar \sigma}_c$, is given by Eq. (\ref {sigmavev}).

\begin{table}[pt]
{\begin{tabular}{@{}c|c|c|c|c|c@{}} 
\hline 
$N$ & ${\bar \sigma}_c$ ($T=\mu=0$) & $T_{c}$ ($\mu =0$) & $\mu _{c}$ ($T=0$) 
& $T_{tcr}$ & 
$\mu _{tcr}$ \\ \colrule
1 &1.78 &  0.96 & 1.24 & 0.46 & 1.06 \\
3 &1.19 & 0.79& 1.07 & 0.24 & 1.034 \\
4 &1.14 & 0.77 & 1.05 & 0.21 & 1.03 \\
6 & 1.09     &  0.75  & 1.03  & 0.19  & 1.02 \\
10 & 1.05 & 0.74 & 1.02 & 0.17 & 1.01
 \\
$\infty $ & 1.0 & 0.72 & 1.0 & 0.0 & 1.0 \\
\hline 
\end{tabular}}
\caption{The relevant physical quantities (in units of $\Lambda$)  
for different values of 
$N$.}
\end{table}

It is also possible to express and analyze the phase diagram in terms of the
pressure $P$ and density $\rho$, defined respectively by $P= - V_{\rm
  eff}({\bar \eta},\bar{\sigma}_c,\mu,T)$ and $\rho = \partial P/\partial
\mu$. This was done in \cite{long}, where it was shown the existence of a
chiral broken/restored coexistence phase (the analog of a liquid-gas phase),
whose result is a direct consequence of the existence of the tricritical point
seen in {}Fig. \ref{phase}.  At this point one may wonder how the
consideration of the order-$\delta^2$ terms, which contains additional $1/N$
as well as new $1/N^2$ corrections, would affect the first order results.
Clearly, at $T \ne 0$ and $\mu \ne 0$ the explicit evaluation of all the three
loop contributions is a rather difficult task. Nevertheless, at $T=0$ and
$\mu=0$, a lengthy evaluation of those contributions has been done in Ref.
\cite{long} showing that the potential corrections to the ${\cal O}(\delta)$
results are not negligible but remain under control, about a few percent.
{}For example, the result obtained for the scalar field vacuum expectation
value at $T=0$ and $\mu=0$ is found to be (at $N=3$) ${\bar \sigma}_c \simeq
1.17 \, \Lambda$ at ${\cal O}(\delta^2)$, for a reasonable choice of the
arbitrary renormalization scale, chosen as given by $\Lambda$, while at ${\cal
  O}(\delta)$ it is ${\bar \sigma}_c \simeq 1.19 \, \Lambda$.  {}For larger
values of $N$ the difference gets even smaller, in conformity with the
large-$N$ results (since at $N\to \infty$ the OPT converges exactly to the
large-$N$ results \cite{LPR}). This then hints on the accuracy and quick
convergence of the order-$\delta$ results discussed in this letter.

\section{Conclusions}

In conclusion, by studying the 2+1 dimensional four-fermion GN model with
discrete chiral symmetry in terms of the OPT, we have shown how this method
can be advantageous compared to the widely used large-$N$ and Monte Carlo
approximations.  Here we have presented precise analytical expressions for the
order parameter ${\bar \sigma}_c$ (at $T=\mu=0$), the critical temperature
$T_c$ (at $\mu=0$), and the critical chemical potential $\mu_c$ (at $T=0$) for
CSR which are valid for any values of $N$ already at lowest order in the OPT.
As usual, all these relations exactly reproduce the standard large-$N$ results
when the $N \to \infty$ limit is taken. We were also able to precisely locate
a tricritical point in the $(T,\mu)$ phase diagram not seen using the
large-$N$ approximation and also not obtained before using other methods,
though numerical simulations seemed to indicate its presence.  We have
discovered that, due to the finite $N$ effects, the model behaves as a Van der
Waals liquid in which a coexistent liquid-gas phase does exist. The observed
metastable region is rather small, which possibly explains why its observation
was missed in Ref.  \cite{lattice2} and previous works.  All these results
seem to survive when going to higher orders in the OPT, as we discussed in the
last section. In this letter we have extended our previous applications of the
OPT to the Gross-Neveu model \cite {LPR,long} by analyzing how the method can
be applied to the Landau expansion for the free energy density in order to
determine the value of critical quantities in a way that is less time
consuming than the straight numerical methods used in Refs. \cite {LPR,long}.
We have shown that this is possible if one considers further approximations in
the optimization equations. However, the numerical effects of such
approximations, even at the first level, turn out to be negligible.

Concerning the next order OPT corrections we have checked in Ref. \cite {long}
that, for $T=0=\mu$, $N=3$, and a reasonable choice of renormalization scale
the differences for the order parameter are about $2\%$. As for the $1/N$
expansion we are not in position to speculate what a complete next to the
leading $1/N^2$ order calculation would give from a quantitatively
perspective. However, we believe that it should point out to the existence of
a tricritical point in the $T-\mu$ plane since, according to our work, this
seems to be an effect of finite $N$ corrections which is also supported by
Monte Carlo applications \cite {lattice1, lattice2}.

Possible follow-ups to this work that would be interesting to investigate
regard the possible existence of a kind of nuclear matter state around
$\mu_c$, at large $\mu$ and low $T$, for which the chiral phase would be
degenerate with a baryonic phase, $\langle \bar{\psi}\psi \rangle \neq 0$ and
$\langle \bar{\psi}\gamma_0 \psi \rangle \neq 0$. This is still an unsettled
question in the literature and that could be ideally addressed with the OPT.
Another interesting work would be to evaluate the various critical (and
tricritical) exponents, determining whether they belong to the mean-field
universality class or the Ising universality class in two dimensions. This is
also a controversial point concerning the GN model, with no clear answer so
far in the literature. Based on its recent success in dealing with critical
theories \cite{LPR,moreOPT,conv} we expect that the method employed here will
be a valuable tool in these analysis providing us with a better understanding
of the chiral symmetry behavior in a hot and dense medium.

{}Finally, we point out that since the GN model is part of a larger class of
four-Fermi models, e.g.  the Nambu-Jona-Lasinio type of models commonly used
in nuclear physics problems and as an effective model for QCD, or even its
non-relativistic version, used in the study of planar superconductors and low
dimensional fermionic systems (like Fermi gases in atomic physics), we expect
that not only our results but also the OPT method used to obtain them can be
of interest and have applications in these areas as well.  In particular, we
believe that the success of the OPT method in redrawing the phase diagram of
the three dimensional GN model could certainly attract the attention of
condensed matter as well as nuclear physicists to this method as an
alternative tool to the well known large-$N$ and Monte Carlo techniques, which
are currently used in an interdisciplinary manner.

\acknowledgments

The authors would like to thank partial support from CNPq, CAPES, UFSC and
FAPERJ (Brazil).

\end{document}